\documentstyle[12pt,epsfig]{article}
\def\shiftdown#1{#1\llap{\lower.04ex\hbox{#1}}}

\begin{document}

\begin{center}

{ \large \em
Current conservation, screening and
the magnetic moment of the $\Delta$ resonance.}
{\footnotemark}

\vspace{5mm}

{ \bf 2.\ Formulation with quark degrees of freedom } 
\vspace{5mm}

{\bf 3.\ Magnetic moment of the $\Delta^o$ and $\Delta^-$ resonances.} 
\footnotetext{ 
Supported by the "Deutsche Forschungsgemeinschaft" under contract
GRK683
}

\end{center}


\vspace{5mm}

\noindent{
{\centerline{\large \bf A.\ I.\ Machavariani$^a$ $^b$ $^c$  and 
Amand Faessler $^a$ } }


}

\vspace{5mm}

\noindent{\small

{  \rm $^a$ 
Institute\ f\"ur\ Theoretische\ Physik\ der\ Univesit\"at\
 T\"ubingen,\newline T\"ubingen\ D-72076, \ Germany}\\

{\rm $^b$ Joint\ Institute\ for\ Nuclear\ Research,\ Dubna,\ Moscow\
region\ 141980,\ Russia}\\

{\rm $^c$ Tbilisi State University, University str.  9,  Tbilisi, Georgia  }\\




}

\vspace{0.5cm}


\medskip
\begin{abstract}

{\bf

Our previous paper \cite{MFNEW} is generalized
within the field theoretical formulation with the quark degrees of
freedom \cite{HW,H,N,Z}, where  
 pions and nucleons are treated as the bound systems of quarks.
It is shown that relations generated by
current conservation for the on shell $\pi N$ bremsstrahlung amplitude
 with composite nucleons and pions
have the same form as 
in the usual quantum field theory \cite{IZ,BD} without quark degrees
of freedom \cite{MFNEW}. Consequently, 
the model independent relations for the  magnetic dipole 
moments of the $\Delta^+$ and $\Delta^{++}$ resonances 
in \cite{MFNEW} remain be the same 
in the quantum field theory with the  
quark degrees of freedom. These relations
are extended for the  magnetic dipole 
moments of the
$\Delta^o$ and  $\Delta^-$ resonances
which are determined  via the anomalous magnetic moment of the neutron
$\mu_n$ as $\mu_{\Delta^o}={ {M_{\Delta}}\over {m_p}} \mu_n$ 
and $\mu_{\Delta^{-}}={3\over 2}\mu_{\Delta^o}$.
}

\end{abstract}


\newpage

\begin{center}
                  {\bf 1. Introduction}
\end{center}
\medskip

The self-consistent generalization of the
conventional quantum field theory for composite 
particles was developed in refs. \cite{HW,H,N,Z}. 
In this approach the quark-hadron bound state functions
 satisfy the appropriate Bethe-Salpeter equations
and  determine the transitions
 from the quark-gluon degrees of freedom into hadron 
degrees of freedom. 
The  general Bethe-Salpeter equations for the quark-hadron
wave functions were derived by Huang and Weldon \cite{HW}.
The basic objects in this approach as well as  in
the Haag-Nishijima-Zimmermann  approaches \cite{H,N,Z}
are the 
creation and annihilation operators of composite particles
which allows  to obtain the $S$-matrix reduction formula
for composite particles. 
The equivalent three dimensional field-theoretical equations for  
composite hadron interaction  amplitudes 
are given in ref. \cite{M1,M2,M3}. The advantages of this
three-dimensional formulation may be 
summarized as follows:  

\begin{itemize}
\item{} 
The poles of the intermediate quark 
and gluon propagators do not contribute to the unitarity condition of
the hadron-hadron scattering amplitudes. Therefore,  
the quark-gluon and hadron degrees of freedom are unambiguously separated and
the problems with the double counting do not appear. 

\item{}
The resulting equations for the hadron-hadron scattering amplitudes
with and without quark degrees of  freedom have the same form.
Therefore, one can easily extend the field-theoretical relations without
quarks to the formulation with the quark-gluon degrees of freedom.

\end{itemize}

The  usual $\pi N$ bremstrahlung amplitude 
$<out;{N'},{\pi'}|{\cal J}^{\mu}(0)|{\pi},N;in>$
with on mass shell  pions and nucleons in the 
$''in''$ and $''out''$ asymptotic
states and with the photon current operator  ${\cal J}^{\mu}(x)$
is depicted by the left diagram in  Fig.1.  In the generalized 
quark-gluon approach \cite{HW,H,N,Z} this amplitude is determined
through the  Green function of 
the transition between the 
$4quark+antiquark$ systems 
and the quark-hadron bound-state wave functions.
The graphical representation 
of the $\pi N$ radiation amplitude
in the quark-gluon approach \cite{HW,H,N,Z} is given by the diagram 
on the right-side  of Fig. 1. 
The construction of  the 
creation and annihilation operators of the asymptotic 
nucleons and pions via  quarks  
 allows to operate with the 
usual expression of the $\pi N$ radiation  
amplitude  $<out;{N'},{\pi'}|{\cal J}^{\mu}(0)|{\pi},N;in>$ because
$$<out;{N'},{\pi'}|{\cal J}^{\mu}(0)|{\pi},N;in>\equiv 
<0|{\cal B}_{out}(N'){\sf a}_{out}(\pi'){\cal J}^{\mu}(0)
{\sf a^+}_{in}(\pi){\cal B^+}_{in}(N)|0>,
\eqno(1.1)$$  
where ${\cal B}_{out(in)}(N)$ and ${\sf a}_{out(in)}(\pi)$ denote the
creation or annihilation operators of the composed nucleon and pion 
in the asymptotic $"out"$ or $"in"$ states.

\vspace{5mm}

\begin{figure}[htb]
\includegraphics[width=13.2cm]{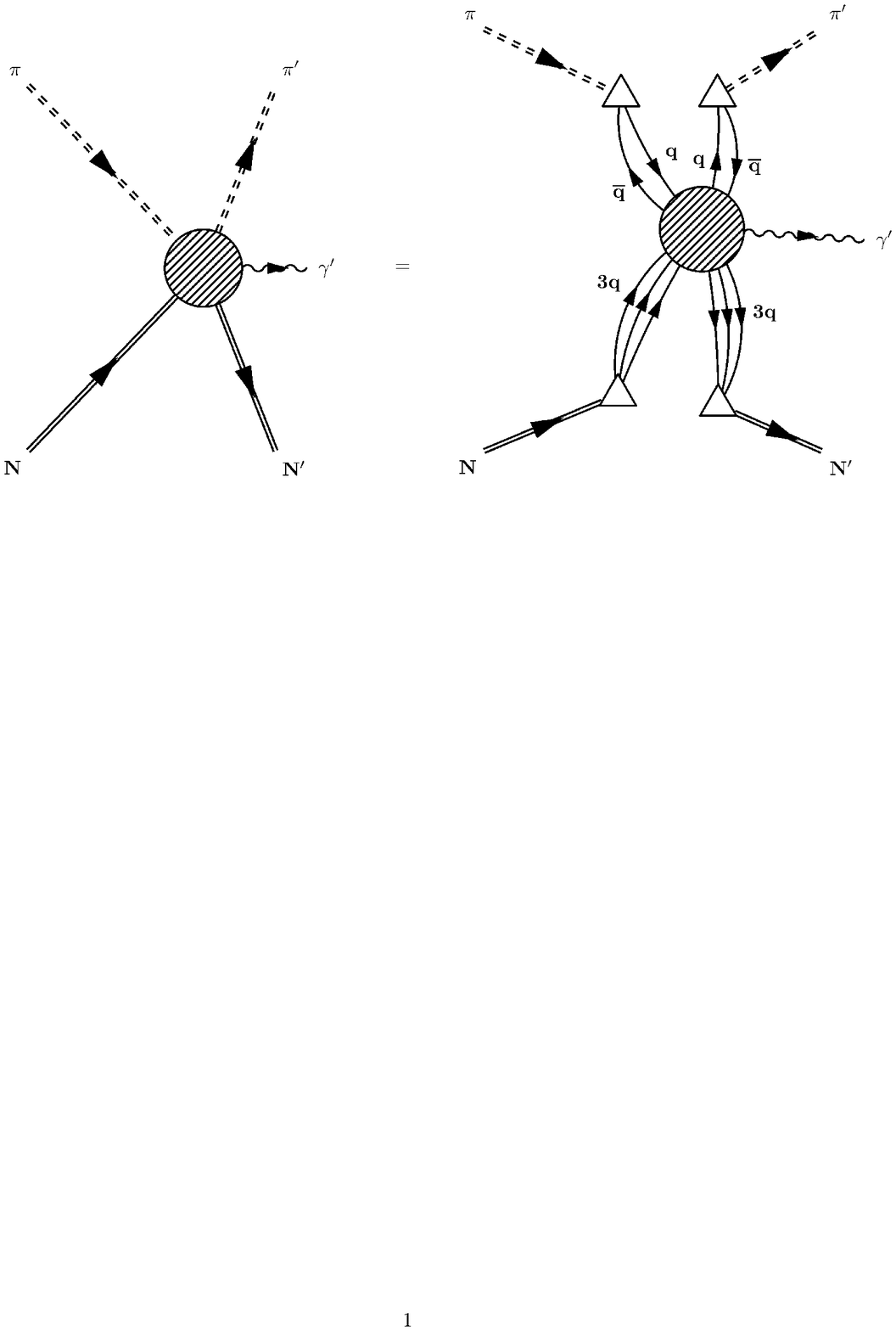}
\vspace{-12.5cm} 
\caption{{\protect\footnotesize {\it
The $\pi N$ bremsstrahlung amplitude  
for composite nucleons and pions.
}} }
\label{fig:one}
\end{figure}

\vspace{5mm}

In this paper  we shall  show that current conservation on the quark level 
for the on shell $\pi N$ bremsstrahlung amplitude (1.1)
takes again the form of the  modified Ward-Takahashi identity
 for the on shell external particle radiation amplitude
${\cal E}^{\mu}_{\gamma'\pi'N'-\pi N}$

$$k'_{\mu}<out;{N'},{\pi'}|{\cal J}^{\mu}(0)|{\pi},N;in>=
k'_{\mu}{\cal E}^{\mu}_{\gamma'\pi'N'-\pi N}+
{\cal B}_{\pi'N'-\pi N}=0,\eqno(1.2a)$$

where $k'_{\mu}$ is the four momentum of the final photon, 
${\cal B}_{\pi'N'-\pi N}$ stands for  a sum
 of the off shell
elastic $\pi N$ scattering amplitudes.

The external particle 
radiation diagrams are depicted in Fig. 2. The only difference between the 
external particle radiation amplitude in Fig. 2 and in the formulation 
without quark degrees of freedom (see Fig. 1 in \cite{MFNEW})
is  the off shell $\pi N$ amplitudes, which contains the nonlocal
momentum-depending source operators of the pion or nucleon.

\vspace{5mm}

\begin{figure}[htb]
\includegraphics[width=14.0cm]{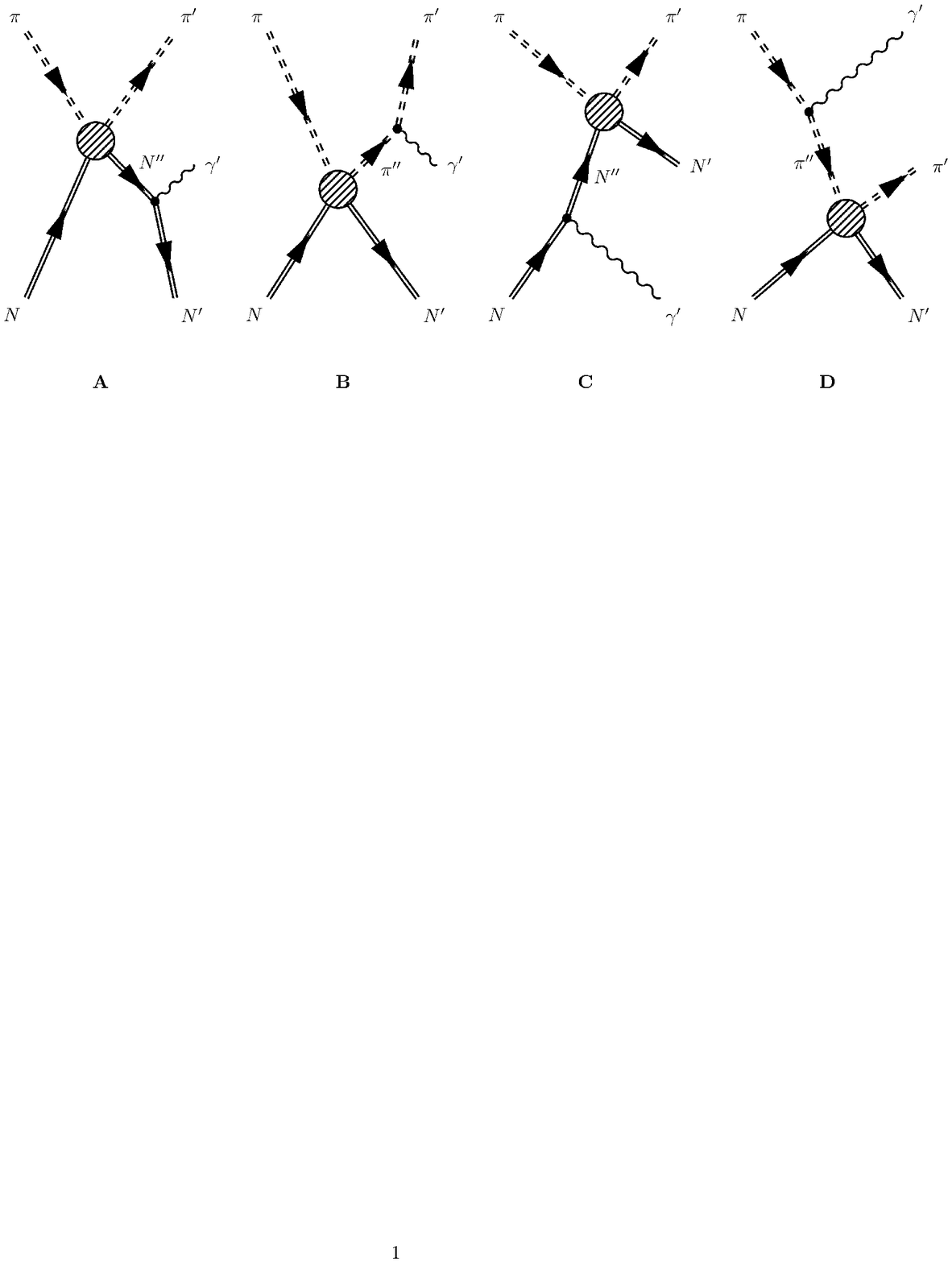}
\vspace{-12.5cm} 
\caption{{\protect\footnotesize {\it
The external particle radiation diagrams of the $\pi N$ bremsstrahlung
amplitude.
}} }
\label{fig:two}
\end{figure}

\vspace{5mm}

Current conservation (1.2a) indicates a connection between the 
four-divergence of the external ${\cal E}^{\mu}_{\gamma'\pi'N'-\pi N}$ 
and internal ${\cal I}^{\mu}_{\gamma'\pi'N'-\pi N}$ particle radiation 
amplitudes ${\cal I}^{\mu}_{\gamma'\pi'N'-\pi N}$

$$k'_{\mu}{\cal E}^{\mu}_{\gamma'\pi'N'-\pi N}
=-k'_{\mu}{\cal I}^{\mu}_{\gamma'\pi'N'-\pi N}=-{\cal B}_{\pi' N'-\pi N},
\eqno(1.2b)$$ 
where 
$<out;{N'},{\pi'}|{\cal J}^{\mu}(0)|{\pi},N;in>=
{\cal E}^{\mu}_{\gamma'\pi'N'-\pi N}
+{\cal I}^{\mu}_{\gamma'\pi'N'-\pi N}$.

\vspace{5mm}

\begin{figure}[htb]
\centerline{\epsfysize=145mm\epsfbox{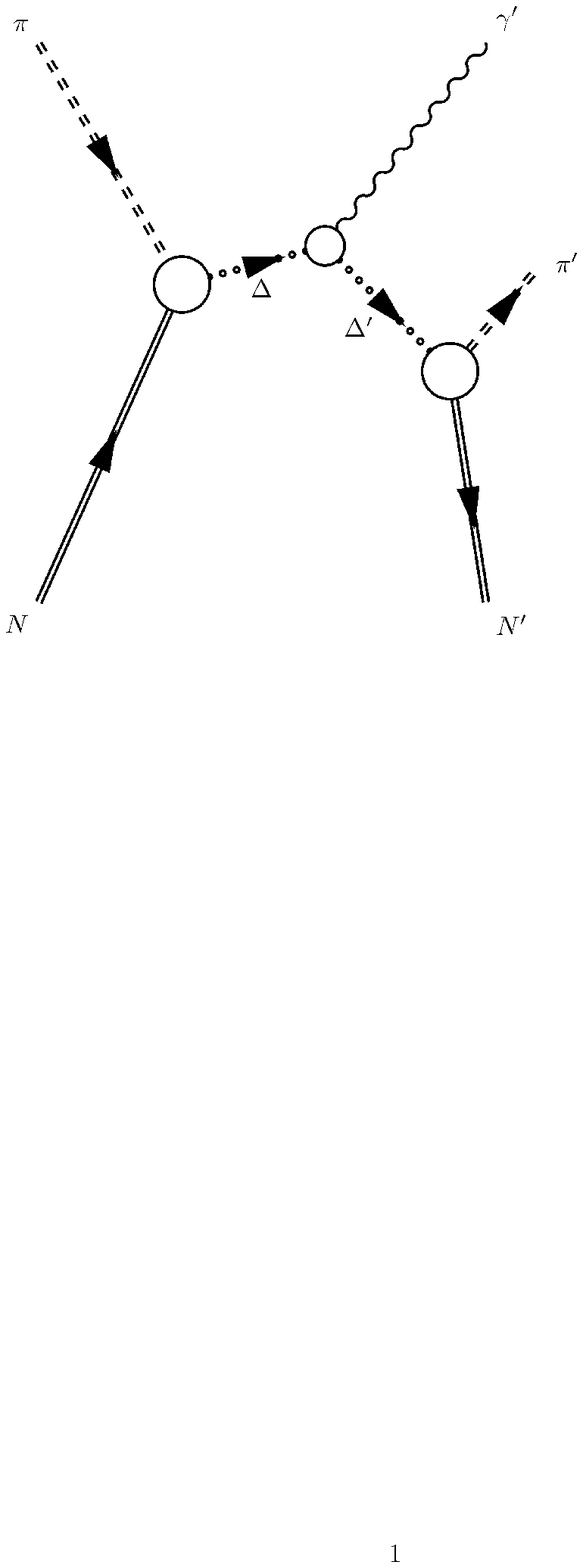}}
\vspace{-8.75cm} 
\caption{{\protect\footnotesize {\it
The double on mass shell $\Delta$ exchange diagram with the 
intermediate $\Delta$ radiation vertex.  
 The $\Delta-\gamma\Delta$ vertex contains the dipole magnetic moment of the 
$\Delta$.}} }
\label{fig:three}
\end{figure}

\vspace{5mm}

One can decompose (1.2a,b) into a set of 
independent current conservation for the 
longitudinal part of the
on shell $\pi N$ radiation amplitude \cite{MFNEW}. 
In particular, one can unambiguously separate the  
$\Delta$-pole parts of the $\pi N$ amplitudes which are contained  in 
${\cal E}^{\mu}_{\gamma'\pi'N'-\pi N}$ and  
${\cal B}_{\pi' N'-\pi N}$. The $\Delta$-pole part of the 
full $\pi N$ Green function determines the $\pi N-\Delta$ wave function  
with the on mass shell $\Delta$ and the effective mass of the $\Delta$ 
\cite{MFNEW,M3,M4,M5}
${\sf m}_{\Delta}(s)=M_{\Delta}(s)-i/2\Gamma_{\Delta}(s)$
which generally depends on the Mandelstam variable $s$.
This $\pi N-\Delta$ wave function and ${\sf m}_{\Delta}(s)$ 
define the intermediate on mass shell $\Delta$ state with the 
four momentum 
$P_{\Delta}=(\sqrt{ {\sf m}^2_{\Delta}(s)+{\bf P}^2_{\Delta}},
{\bf P}_{\Delta})$ also in the present formulation with quark
degrees of freedom. The considered field theoretical  
 $\pi N$ bremsstrahlung amplitudes are not depending on the model of
${\sf m}_{\Delta}(s)$ which must be determined  separately.  
The sum   of the $\Delta$-pole parts of the off shell $\pi N$ amplitudes in 
${\cal E}^{\mu}_{\gamma'\pi'N'-\pi N}$ (Fig. 2) 
 reproduces the double on mass shell 
$\Delta$ exchange amplitude
$({\cal E_L}^{3/2})^{\mu}_{\gamma'\pi'N'-\pi N}(\Delta-\gamma\Delta)$ 
which contains the $\Delta-\gamma\Delta$ 
vertex with the 
anomalous magnetic moment of the proton instead of the  magnetic dipole moment 
of the $\Delta$. 
$({\cal E_L}^{3/2})^{\mu}_{\gamma'\pi'N'-\pi N}(\Delta-\gamma\Delta)$ has the 
same analytical structure as the intermediate $\Delta$ radiation diagram 
${\cal I}^{\mu}_{\gamma'\pi'N'-\pi N}(\Delta-\gamma\Delta)$ in Fig. 3.
This amplitude is unambiguously separated
from the internal particle radiation diagrams 
based on the $\Delta$-pole terms of the intermediate $\pi N$ 
Green function \cite{MFNEW}. Thus
$$k'_{\mu}({{\cal E_L}^{3/2}})^{\mu}_{\gamma'\pi'N'-\pi N}(\Delta-\gamma\Delta)
=-{k'}_{\mu}{\cal I}_{\gamma'\pi' N'-\pi N}^{\mu}(\Delta-\gamma\Delta)=
-{\cal B}^{3/2}_{\pi' N'-\pi N}(\Delta-\gamma\Delta),
\eqno(1.3)$$     
where the lower index $ _{\cal L}$ and the upper index $ ^{3/2}$ denotes 
the longitudinal and the spin-isospin  $(3/2,3/2)$ part of the corresponding 
expressions. 
The identical structure  of \newline   
$({\cal E_L}^{3/2})_{\gamma'\pi' N'-\pi N}^{\mu}({\Delta}-\gamma\Delta)$ 
and 
${\cal I}_{\gamma'\pi' N'-\pi N}^{\mu}({\Delta}-\gamma\Delta)$
in Fig. 3 allows to obtain an analytical and model independent relations
between the  magnetic dipole moments of the $\Delta^+$ and $\Delta^{++}$ 
resonances and the anomalous magnetic moment of the proton. 
In this paper we
generalize this relation for the  magnetic  dipole moments of the 
$\Delta^o$ and $\Delta^{-}$ resonances.

This paper consists of four Sections. 
The creation and annihilation operators of  composite particle
and corresponding Ward-Takahashi identity are constructed
in  the next Section.
The model-independent relation
between  the  magnetic dipole moments of the 
$\Delta^o$ and 
$\Delta^{-}$ resonances and the anomalous magnetic moment of the neutron
is given in Section 3.
The conclusions and comparison of the suggested relations
for the magnetic  dipole moments of 
${\Delta^o}$ and ${\Delta^-}$ with the numerical values of
other authors are presented in Sect. 4.


\vspace{0.5cm}

\begin{center}
                  {\bf 2. The Ward-Takahashi identity for the on shell
$\pi N$ bremstrahlung amplitude
in  the field theoretical approach with the quark degrees of freedom}
\end{center}

\vspace{0.25cm}

\par

The creation and annihilation operators of the hadrons
as the quark  cluster operators were constructed in ref. \cite{HW}
in the framework of the usual  quantum field theory. 
The corresponding nucleon and pion field operators 
$\Psi_{p_N}(Y)$ and $\Phi_{p_{\pi}}(X)$
are composed through the  local quark
field operators $q(x)$. $\Psi_{p_N}(Y)$ and $\Phi_{p_{\pi}}(X)$  
are nonlocal because they depend on the nucleon and pion four moments 
$p_N$ and $p_{\pi}$ correspondingly

$$\Psi_{\bf p_N}(Y)=
\int d^4r_3d^4r_{1,2}{\widetilde \chi}^{\dag}_{\bf p_N}(Y=0,r_{1,2}.r_3)
{\sf T}\biggl( q_1(y_1)q_2(y_2)q_3(y_3) \biggr),\eqno(2.1)$$

where $Y$, $r_{1,2}$ and $r_3$ are the Jacobi coordinates  
$y_1=Y-\eta_3r_3+\eta_2r_{1,2}$, $y_2=Y-\eta_3r_3-\eta_1r_{1,2}$, 
 $y_3=Y+\eta_{1,2}r_3$
with $\eta_3+\eta_{1,2}=1$ and $\eta_1+\eta_2=1$,

$$\chi_{\bf p_N}(y_1,y_2,y_3)=
<0|{\sf T}\biggl(q_i(y_1)q_j(y_2)q_k(y_3)|{\bf p}_N, s_N,i_N;in>=
 e^{-ip_NY}\chi_{\bf p_N}(Y=0,r_{1,2},r_3)\eqno(2.2a)$$
is a solution of the Bethe-Salpeter equation for the  three quark bound 
state with the nucleon mass $m_N$ and four momentum 
$p_N=(\sqrt{{\bf p_N}^2+m_N^2},{\bf p_N})$; $s_N$ and $i_N$ denotes the
 spin-isospin projections of the nucleon.
 ${\widetilde \chi}^{\dag}_p$ satisfies the normalization condition \cite{IZ}

$${1\over{2im_N^2}}<\chi_{\bf p'_N}|p_N^{\mu}{{\partial}\over
{\partial p_N^{\mu}}}
\Bigl[G^{-1}(3q)\Bigr]|\chi_{\bf p_N}>\equiv
<{\widetilde \chi}_{\bf p'_N}|\chi_{\bf p_N}>
=<{\bf p'_N},s'_N,i'_N;m_N|{\bf p_N},s_N,i_N;m_N>,\eqno(2.2b)$$
where $G(3q)$ is the full Green function of three interacting 
quarks{\footnotemark}.

\footnotetext{
The on mass shell $\Delta$ state with the complex mass 
$m_{\Delta}=M_{\Delta}+i\Gamma_{\Delta}/2$
can be constructed through the intermediate three quark state 
in the same manner as the one nucleon state. 
 In particular, it is necessary to find the solution 
of the Bethe-Salpeter equation for the 
 $\Delta$-pole state of the $3quark-3quark$ Green function:
$\Psi_{{\bf p}_{\Delta}}(x_1,x_2,x_3)=
<0|T\bigl(q_1(x_1)q_2(x_2)q_3(x_3)\bigr)|\Psi_{{\bf p}_{\Delta}}>$.}

The asymptotic nucleon annihilation (creation) 
operator ${\cal B}^{in(out)}({\bf p}_N)$
for an on mass-shell nucleon is determined as

$${\cal B}_{in(out)}({\bf p}_N)=\lim_{x^o\to -\infty(+\infty)}
{\cal B}_{\bf p_N}(x^o),\eqno(2.3a)$$
where the weak limit $\lim_{x^o\to -\infty(+\infty)}$ is assumed. 
The Heisenberg operator ${\cal B}_{\bf p_N}(X^o)$ is given in the same
form as in  conventional quantum field theory
$${\cal B}_{\bf p_N}(x^o)=\int d^3{\bf x}\exp{(ip_Nx)}{\overline u}({\bf p}_N)
\gamma_o\Psi_{\bf p_N}(x).\eqno(2.3b)$$

The composite meson fields are constructed  
using the quark-antiquark operator

$$\Phi_{{\bf p}_{\pi}}(X)=\int d^4\rho_{1,2}
{\widetilde \phi}^+_{\bf p_{\pi}}(X=0,\rho_{1,2}){\sf T}\biggl(q_i(x_1)
{\overline q}_i(x_2)\biggr),\eqno(2.4a)$$

where $p_{\pi}=(\sqrt{{\bf p}_{\pi}^2+m_{\pi}^2},{\bf p}_{\pi})$,
$x_1=X+\mu_2 \rho_{1,2}$, 
$x_2=X-\mu_1 \rho_{1,2}$ with $\mu_1+\mu_2=1$  and
$$\phi_{\bf p_{\pi}}(x,y)=
<0|{\sf T}\biggl(q_i(x){\overline q}_i(y)
\biggr)|{\bf p_{\pi}},i_{\pi};m_{\pi}>\eqno(2.4b)$$
is the solution of the Bethe-Salpeter equation 
of the quark-antiquark bound state. This function satisfies 
the normalization condition  

$${!\over{2im_{\pi}^2}}<\phi_{{\bf p'}_{\pi}}|p^{\mu}_{\pi}
{{\partial}\over{\partial p^{\mu}_{\pi}}}
g^{-1}(q{\overline q})|\phi_{{\bf p}_{\pi}}>\equiv
<{\widetilde \phi}_{{\bf p'}_{\pi}}|\phi_{{\bf p}_{\pi}}>
=<{\bf p'_{\pi}},i'_{\pi};m_{\pi}|{\bf p_{\pi}},i_{\pi};m_{\pi}>
,\eqno(2.4c)$$
where $g(q{\overline q})$ is the full quark-antiquark Green function. 

The asymptotic meson creation or annihilation operator is 

$${\sf a}_{in(out)}({\bf p}_{\pi})=\lim_{x^o\to -\infty(+\infty)}
{\sf a}_{{\bf p}_{\pi}}(x^o),\eqno(2.5a)$$
where 
$${\sf a}_{{\bf p}_{\pi}}(x^o)=\int d^3{\bf x}\exp{(ip_{\pi}x)}\Bigl[
{{\partial}\over{\partial x^o}}-ip_{\pi}^o\Bigr]\Phi_{p_{\pi}}(x).\eqno(2.5b)$$

The composite operators (2.3a) and (2.5a) 
satisfy the same commutation relations as the ordinary  local field
operators of the asymptotic nucleons and pions in the usual 
quantum field theory ~\cite{IZ,BD}

$$
\biggl\{{\cal B}_{in(out)}^+({\bf p'}),
{\cal B}_{in(out)}({\bf p})\biggl\}=
(2\pi)^3{{p_N^o}\over{m_N}}\delta({\bf p'-p});$$
$$\biggl\{{\cal B}_{in(out)}({\bf p'}),
{\cal B}_{in(out)}({\bf p})\biggr\}=
\biggl\{ {\cal B}_{in(out)}^+({\bf p'}),
{\cal B}^+_{in(out)}({\bf p})\biggr\}=0,\eqno(2.6a)$$

$$\biggl[{\sf a}_{in(out)}^+({\bf p'}_{\pi}),
{\sf a}_{in(out)}({\bf p}_{\pi})\biggr]=(2\pi)^32p^o_{\pi}
\delta({\bf p'_{\pi}-p_{\pi}});$$
$$\biggl[{\sf a}_{in(out)}({\bf p'}_{\pi}),
{\sf a}_{in(out)}({\bf p}_{\pi})\biggr]=
\biggl[{\sf a}_{in(out)}^+({\bf p'}_{\pi}),
{\sf a}_{in(out)}^+({\bf p}_{\pi})\biggr]=0
.\eqno(2.6b)$$

The relations (2.6a,b)
allow to build  any $"in"$ or $"out"$ hadron states
through the intermediate quark-cluster states. 
These operators form the usual completeness condition
for the asymptotic $"in"$ or $"out"$ hadron fields
$\sum_n |n;in(out)>$ $<(out)in;n|={\widehat{\bf 1}}$ and
the well-known  ${\cal S}$-matrix element as ${\cal S}_{nm}=<out; n|m;in>$.
In contrast to  
local quantum field theory, the Heisenberg fields ${\cal B}_{\bf p}(x^o)$
(2.3b) and ${\sf a}_{\bf p}(x^o)$ (2.5b)
do not satisfy the equal-time commutation relations (2.6a,b)
$\biggl\{{\cal B}_{\bf p'}(x_o)$,
 ${\cal B}_{\bf p}(x_o)\biggr\}\ne 0$,
 $\biggl[ {\cal B}_{\bf p'}(x_o),
{\sf a}_{{\bf p}_{\pi}}(x_o)\biggr]\ne 0$, etc.
Nevertheless, 
the basic relations of the usual quantum field theory
remain the same 
in the field theoretical approach with quarks. In particular,
for the $\pi N$ radiation amplitude $A_{\gamma'\pi' N'-\pi N}^{\mu}$
one has 

$$A_{\gamma'\pi' N'-\pi N}^{\mu}
=i\int d^4ze^{ik'z}<0|
{\cal B}_{out}({\bf p'}_N){\sf a}_{out}({\bf p'}_{\pi}){\cal J}^{\mu}(z)
{\sf a^+}_{in}({\bf p}_{\pi}){\cal B^+}_{in}({\bf p}_N)|0>,\eqno(2.7a)$$
where $k'_{\mu}$ denotes the four momentum of the on shell final photon
$k'_{\mu}{k'}^{\mu}=0$ and 
$k'_{\mu}=(p_N+p_{\pi}-p'_N-p'_{\pi})_{\mu}\equiv(P-P')_{\mu}$;
$${\cal J}^{\mu}(z)={\overline q}(z)
\Bigl( {{\lambda^3}\over{2}}+{{\lambda^8}\over{2\sqrt{3}}}\Bigr)\gamma^{\mu}
q(z)\eqno(2.7b)$$ 
denotes the photon source operator with the Gell-Mann 
flavor matrices $\lambda$ \cite{IZ}. 

A symbolical picture 
of the $\pi N$ bremsstrahlung amplitude 
(2.7a) with the intermediate  quark-clusters states is given in Fig. 1.
The triangles in Fig. 1 describe the quark-hadron bound state wave functions   
(2.2a), (2.4a) and their orthogonal expressions.
These vertices play the role of the hadronization
functions.   
Consequently, the  $\pi N\Longrightarrow \gamma'\pi'N'$ 
amplitude (2.7a) is replaced  by the 
$4q{\overline q}-\gamma'4q'{\overline q}'$ transition amplitude.
Using the generalized ${\cal S}$-matrix reduction formula \cite{HW}
we obtain

$${k'}_{\mu}A_{\gamma'\pi' N'-\pi N}^{\mu}={\overline u}({\bf p'}_N)
(\gamma_{\nu} {p'_N}^{\nu}-m_N)({p'_{\pi}}^2-m_{\pi}^2)
\ {k'}_{\mu}{\sc G}^{\mu}\ 
(\gamma_{\nu} {p_N}^{\nu}-m_N)({p_{\pi}}^2-m_{\pi}^2)
 u({\bf p}_N),\eqno(2.8a)$$

where

$${k'}_{\mu}{\sc G}^{\mu}=i\int d^4y'_1d^4y'_2d^4y'_3
d^4x'_1d^4x'_2d^4y_1d^4y_2d^4y_3d^4x_1d^4x_2
e^{ik'z}d^4z
{\widetilde \chi}^+_{\bf p'_N}(y'_1,y'_2,y'_3)
{\widetilde \phi}^+_{\bf p'_{\pi}}(x'_1,x'_2)
{{\partial}\over{\partial z^{\mu}} }$$
$$<0|{\sf T}\biggl(
{\sf T}\Bigl(q_1(y'_1)q_2(y'_2)q_3(y'_3)\Bigr)
{\sf T}\Bigl(q_1(x'_1){\overline q}_2(x'_2)\Bigr)
{\cal J}^{\mu}(z)
{\sf T}\Bigl({\overline q}_1(y_1){\overline q}_2(y_2){\overline q}_3(y_3)\Bigr)
{\sf T}\Bigl({\overline q}_1(x_1)q_2(x_2)\Bigr)\biggr)|0>$$
$$\chi_{\bf p_N}(y_1,y_2,y_3)\phi_{\bf p_{\pi}}(x_1,x_2),\eqno(2.8b)$$
where the double time-ordered product is defined as   
${\sf T}\biggl({\sf T}\Bigl({\overline q}_1(x_1)q_2(x_2)\Bigr)
{\cal J}^{\mu}(z)\biggr)=$

$\Bigl({\overline q}_1(x_1)\theta(x_1^o-x_2^o)q_2(x_2)
\theta(x_2^o-z^o){\cal J}^{\mu}(z)-
{\overline q}_2(x_2)\theta(x_2^o-x_1^o)q_1(x_1)
\theta(x_1^o-z^o){\cal J}^{\mu}(z)\Bigr)$

$+
\Bigl({\cal J}^{\mu}(z)\theta(z^o-x_1^o)
{\overline q}_1(x_1)\theta(x_1^o-x_2^o)q_2(x_2)-
{\cal J}^{\mu}(z)\theta(z^o-x_2^o)
{\overline q}_2(x_2)\theta(x_2^o-x_1^o)q_1(x_1)
\Bigr)$. 

The choice of the 
average $X=x_1+x_2$, $Y=y_1+y_2+y_3$
or the c.m. coordinates
$X=\mu_1x_1+\mu_2x_2$, $\rho_{1,2}=x_1-x_2$ and 
$Y=\eta_3y_3+\eta_{1,2}Y_{1,2}$, $\rho_3=y_3-Y_{1,2}$, 
$Y_{1,2}=\eta_1y_1+\eta_2y_2$ in (2.1) and in (2.4a) is not unique. 
But the $S$-matrix
reduction formula and (2.8a,b) are not depend on the choice of 
$X$ and $Y$.

The important property of the on shell 
amplitude (2.8a) is that only the operators 
${\overline u}({\bf p}_N)(\gamma_{\nu} {\bf p_N}^{\nu}-m_N) 
{\widehat {\sc P}}_{N'}\equiv
{\overline u}({\bf p}_N)(\gamma_{\nu} {p_N}^{\nu}-m_N)\int d^4y_1d^4y_2d^4y_3 
{\widetilde \chi}^+_{\bf p_N}(y_1,y_2,y_3)
{\sf T}\Bigl(q_1(y_1)q_2(y_2)q_3(y_3)\Bigr)$, 
$({p_{\pi}}^2-m_{\pi}^2)
{\widehat {\sc P}}_{\pi}\equiv({p_{\pi}}^2-m_{\pi}^2)\int d^4x_1d^4x_2
{\widetilde \phi}^+_{\bf p_{\pi}}(x_1,x_2 )
{\sf T}\Bigl(q_1(x_1){\overline q}_2(x_2)\Bigr)$
and their Hermitian conjugate produce the asymptotic one-nucleon 
and one-pion states. Therefore, the zeros of  the Dirac
$(\gamma_{\nu} {p'_N}^{\nu}-m_N)$, $(\gamma_{\nu} {p_N}^{\nu}-m_N)$ 
and the Klein-Gordon operators 
$({p'_{\pi}}^2-m_{\pi}^2)$, $({p_{\pi}}^2-m_{\pi}^2)$ 
in the on shell amplitude (2.8a) 
can be compensated only by  
${\widehat {\sc P}}_{N'}$, ${\widehat {\sc P}}_{\pi'}$ and their Hermitian
conjugate operators. But 
${\widehat {\sc P}}_{N'}$, ${\widehat {\sc P}}_{\pi'}$ and their 
conjugate are included in  ${\sc G}^{\mu}$ (2.8b).
The remaining part  of the full Green function 
$${\tau}^{\mu}={\sc G}^{\mu}+{\em g}^{\mu}=$$
$$\int d^4y'_1d^4y'_2d^4y'_3
d^4x'_1d^4x'_2d^4y_1d^4y_2d^4y_3d^4x_1d^4x_2
e^{ik'z}d^4z
{\widetilde \chi}^+_{\bf p'_N}(y'_1,y'_2,y'_3)
{\widetilde \phi}^+_{\bf p'_{\pi}}(x'_1,x'_2)$$
$$ <0|{\sf T}\biggl(
q_1(y'_1)q_2(y'_2)q_3(y'_3)
q_1(x'_1){\overline q}_2(x'_2)
{\cal J}^{\mu}(z)
{\overline q}_1(y_1){\overline q}_2(y_2){\overline q}_3(y_3)
{\overline q}_1(x_1)q_2(x_2)\biggr)|0>$$
$$\chi_{\bf p_N}(y_1,y_2,y_3)\phi_{\bf p_{\pi}}(x_1,x_2)\eqno(2.8c)$$
involves all possible exchanges of the quark operators 
between ${\sf T}\Bigl(q_1(y'_1)q_2(y'_2)q_3(y'_3)\Bigr)$,
${\sf T}\Bigl(q_1(x'_1){\overline q}_2(x'_2)\Bigr)$,
${\sf T}\Bigl(q_1(y_1)q_2(y_2)q_3(y_3)\Bigr)$ and 
${\sf T}\Bigl(q_1(x_1){\overline q}_2(x_2)\Bigr)$.
Therefore,  ${\em g}^{\mu}$
does not contribute
into the on shell amplitude (2.7a), because  ${\em g}^{\mu}$ 
does not contain
${\widehat {\sc P}}_{N'}$, ${\widehat {\sc P}}_{N}$, 
${\widehat {\sc P}}_{\pi'}$, ${\widehat {\sc P}}_{\pi'}$.\footnotemark
 \footnotetext{
This can be also verified 
using the limits over the
coordinates
 ${ {X'}^o}\Longrightarrow \infty$, ${ {X}^o}\Longrightarrow -\infty$
 ${ {Y'}^o}\Longrightarrow \infty$, ${ {Y}^o}\Longrightarrow -\infty$
in the quark 
and composite particle  operators (2.3a,b), (2.5a,b).} 
Correspondingly, one can replace
${\sc G}^{\mu}$ (2.8b) with ${\tau}^{\mu}$ (2.8c) in (2.8a)

$${k'}_{\mu}A_{\gamma'\pi' N'-\pi N}^{\mu}={\overline u}({\bf p'}_N)
(\gamma_{\nu} {p'_N}^{\nu}-m_N)({p'_{\pi}}^2-m_{\pi}^2)
\ {k'}_{\mu}{\tau}^{\mu}\
(\gamma_{\nu} {p_N}^{\nu}-m_N)({p_{\pi}}^2-m_{\pi}^2)
 u({\bf p}_N),\eqno(2.8d)$$

Based on the equal-time commutation relations between the 
quark operators it is easy to get
the equal-time commutation rules for the 
photon source operator (2.7b) and the 
quark operators

$$\biggl[ {\cal J}^{o}(z),q_j(y')
\biggr]\delta(z_o-y'_o)=-e_{j}\delta^{(4)}(z-y')q_j(y');\ \ \
\biggl[ {\cal J}^{o}(z),{\overline q}_J(y)
\biggr]\delta(z_o-y_o)=e_J\delta^{(4)}(z-y){\overline q}_J(y),\eqno(2.9)$$
where $e_j$ denotes the charge of the quark $j$.
In particular, $e_j=2/3\ e$ for the $u$-quark and $e_j=-1/3\ e$ for
the $d$ quarks.
The equal-time commutators (2.9) and  integration over $z$ 
in (2.8c) allows to rewrite  (2.8d) as

$${k'}_{\mu}A_{\gamma'\pi' N'-\pi N}^{\mu}=
-i{\overline u}({\bf p'}_N)(\gamma_{\nu} {p'_N}^{\nu}-m_N)
\int d^4y'_1d^4y'_2d^4y'_3{\widetilde \chi}^+_{\bf p'_N}(y'_1,y'_2,y'_3)$$
$$\Bigl[e_1^{N'}e^{ik'y'_1}+e_2^{N'}e^{ik'y'_2}
+e_3^{N'}e^{ik'y'_3}\Bigr]<out;{\bf p'}_{\pi}|
{\sf T}\Bigl(q_1(y'_1)q_2(y'_2)q_3(y'_3)\Bigr)
|{\bf p}_{\pi}{\bf p}_N;in>$$
$$-i({p'_{\pi}}^2-m_{\pi}^2)\int d^4x'_1d^4x'_2
{\widetilde \phi}^+_{\bf p'_{\pi}}(x'_1,x'_2)
\Bigl[e_1^{\pi'}e^{ik'x'_1}+e_2^{\pi'}e^{ik'x'_2}\Bigr]
<out;{\bf p'}_{N}|{\sf T}\Bigl(q_1(x'_1){\overline q}_2(x'_2)\Bigr)
|{\bf p}_{\pi}{\bf p}_N;in>$$
$$+i\int d^4y_1d^4y_2d^4y_3
\Bigl[e_1^{N}e^{ik'y _1}+e_2^{N}e^{ik'y_2}+e_3^{N}e^{ik'y_3}\Bigr]
<out;{\bf p'}_{N}{\bf p'}_{\pi}
|{\sf T}\Bigl({\overline q}_1(y_1){\overline q}_2(y_2){\overline q}_3(y_3)
\Bigr)|{\bf p}_{\pi};in>$$
$$\chi_{\bf p_N}(y_1,y_2,y_3)
(\gamma_{\nu} {p_N}^{\nu}-m_N)({p_{\pi}}^2-m_{\pi}^2)u({\bf p}_N)$$
$$+i({p'_{\pi}}^2-m_{\pi}^2)\int d^4x_1d^4x_2
\Bigl[e_1^{\pi}e^{ik'x_1}+e_2^{\pi}e^{ik'x_2}\Bigr]
<out;{\bf p'}_{N}{\bf p'}_{\pi}|{\sf T}\Bigl( {\overline q}_1(x_1)q_2(x_2)
\Bigl)|{\bf p}_{N};in>\phi_{\bf p_{\pi}}(x_1,x_2)
.\eqno(2.11)$$

After integration over the coordinates $X'$, $X$, $Y'$ and $Y$
 we obtain

$${k'}_{\mu}A_{\gamma'\pi' N'-\pi N}^{\mu}
({\bf p'_{\pi},p'_N,k';p_{\pi},p_N})=-i
(2\pi)^4\ \delta^{(4)}(p'_N+p'_{\pi}+k'-p_{\pi}-p_N)$$
$$\Biggl[{\overline u}({\bf p'}_N)(\gamma_{\nu} {p'_N}^{\nu}-m_N)
{ {e_{N'}}\over{\gamma_{\nu} (p'_N+k')^{\nu}-m_N  }}
<out;{\bf p'}_{\pi}|J_{\bf p'_N,k'}(0)|{\bf p}_{\pi}{\bf p}_N;in>$$
$$+( {p'}_{\pi}^2-m_{\pi}^2){{e_{\pi'}}\over{(p'_{\pi}+k')^2-m_{\pi}^2} }
<out;{\bf p'}_{N}|j_{\bf p'_{\pi},k'}(0)|{\bf p}_{\pi}{\bf p}_N;in>$$
$$-
<out;{\bf p'}_{\pi}{\bf p'}_{N}|{\overline J}_{\bf p_N,k'}(0)|{\bf p}_{\pi};in>
{{e_{N}}\over{\gamma_{\nu} (p_N-k')^{\nu}-m_N} }(\gamma_{\nu} {p_N}^{\nu}-m_N)
u({\bf p}_N)$$
$$-<out;{\bf p'}_{\pi}{\bf p'}_{N}|j_{\bf p_{\pi},k'}(0)|{\bf p}_{N};in>
{{e_{\pi}}\over{(p_{\pi}-k')^2-m_{\pi}^2}}
(p_{\pi}^2-m_{\pi}^2)\Biggr],\eqno(2.12)$$

where the source operators of the nucleon and pion are constructed 
through the nonlocal nucleon and pion fields  (2.1) and (2.4a)
as

$$J_{\bf p'_{N},k'}(Y)=
\bigl(i\gamma_{\sigma}\nabla_Y^{\sigma}-m_{N}\bigr)\Psi_{{\bf p}_{N},k'}(Y)
;\ \ \Psi_{{\bf p'}_{N},k'}(Y')=
\int d^4r'_3d^4r'_{1,2}
{\widetilde \chi}^+_{\bf p'_N}(Y'=0,r'_{1.2},r'_3)$$
$$\Bigl[{{e_1^{N'}}\over {e_N'}}e^{ik'(-\eta_3r'_3+\eta_2r'_{1,2})}+
{{e_2^{N'}}\over {e_N'}}e^{-ik'(\eta_3r'_3+\eta_1r'_{1,2})}
+{{e_3^{N'}}\over {e_N'}}e^{ik'\eta_{1,2}r'_3}\Bigr]
{\sf T}\Bigl(q_1(y'_1)q_2(y'_2)q_3(y'_3)\Bigr)\eqno(2.13a)$$

$$ j_{\bf { p'}_{\pi},k'}(X)=
\bigl(\Box_X+m_{\pi}^2\bigr)\phi_{ p_{\pi}}(X)
;\ \ \phi_{\bf p_{\pi},k'}(X')=
\int d^4\rho'_{1,2}
{\widetilde \phi}^+_{\bf p'_{\pi}}(X'=0,\rho'_{1,2})$$
$$\Bigl[{{e_1^{\pi'}}\over {e_{\pi'}}}e^{ik'\mu_2\rho'_{1,2}}+
{{e_2^{\pi'}}\over {e_{\pi'}}}e^{-ik'\mu_1\rho'_{1,2}}\Bigr]
{\sf T}\Bigl(q_1(x'_1){\overline q}_2(x'_2)\Bigr).
\eqno(2.13b)$$

The charge factors in front of the off shell $\pi N$ amplitudes 
in (2.12) are extracted in analogue to the formulation without quark 
degrees of freedom \cite{MFNEW}, where  
the charge of the nucleon and 
pion arise  in the equal time commutators due to charge conservation. 
In the source operators (2.13a,b)
the quark charges are distributed according to the  
commutators (2.9). 
This distribution is a result of the choice of the field operators 
(2.1) and (2.4a).
Other  choices of the source operators of the composite nucleons and
pions are considered in Appendix A.
Unlike (2.13a,b) other source operators  do not contain 
the quark charge distributions.


Because of the
zeros of  the free Dirac
and the Klein-Gordon operators
$(\gamma_{\nu} {p'_N}^{\nu}-m_N)$, $(\gamma_{\nu} {p_N}^{\nu}-m_N)$, 
$({p'_{\pi}}^2-m_{\pi}^2)$, $({p_{\pi}}^2-m_{\pi}^2)$ equation
(2.12) corresponds to current conservation
 ${k'}_{\mu}A_{\gamma'\pi' N'-\pi N}^{\mu}=0$ for any $k'$. 
For  $k'=0$ ${k'}_{\mu}A_{\gamma'\pi' N'-\pi N}^{\mu}=0$  according to 
cancellations of the on shell $\pi N$ amplitudes.
Therefore, (2.12) represents current conservation 
for the on-mass shell $\pi  N$ bremsstrahlung amplitude

$${k'}_{\mu}
\Biggl[ A_{\gamma'\pi' N'-\pi N}^{\mu}
({\bf p'_{\pi},p'_N,k';p_{\pi},p_N})
\Biggr]_{on\ mass\ shell\ \pi',\ N',\ \pi,\ N}=0.\eqno(2.14)$$

Following \cite{MFNEW} we extract the full energy-momentum conservation 
$\delta$ function from the  radiative $\pi N$ scattering 
amplitude $A_{\gamma'\pi' N'-\pi N}^{\mu}$ and introduce the corresponding
non-singular amplitude
$<out;{\bf p'}_{N}{\bf p'}_{\pi}|{\cal J}^{\mu}(0)|{\bf p}_{\pi}{\bf p}_N;in>$

$${k'}_{\mu}A_{\gamma'\pi' N'-\pi N}^{\mu}=-i
(2\pi)^4\ \delta^{(4)}(p'_N+p'_{\pi}+k'-p_{\pi}-p_N)
{k'}_{\mu}
<out;{\bf p'}_{N}{\bf p'}_{\pi}|{\cal J}^{\mu}(0)|{\bf p}_{\pi}{\bf p}_N;in>.
\eqno(2.15)$$

The identity $a/(a+b)\equiv 1- b/(a+b)$ and (2.15)
allows to rewrite  (2.12) as

$$
{k'}_{\mu}
<out;{\bf p'}_{N}{\bf p}_{\pi'}|{\cal J}^{\mu}(0)|{\bf p}_{\pi}{\bf p}_N;in>=
{\cal B}_{\pi' N'-\pi N}+{k'}_{\mu}{\cal E}_{\gamma'\pi' N'-\pi N}^{\mu}
=0,\eqno(2.16)$$

where

$${\cal B}_{\pi' N'-\pi N}=
e_{N'}{\overline u}({\bf p'}_N)
<out;{\bf p'}_{\pi}|J_{\bf p'_{N}k'}(0)|{\bf p}_{\pi}{\bf p}_N;in>
+e_{\pi'}<out;{\bf p'}_{N}|j_{\bf { p'}_{\pi}k'}(0)|{\bf p}_{\pi}{\bf p}_N;in>$$
$$-e_N<out;{\bf p'}_{\pi}{\bf p'}_{N}|{\overline J}_{\bf p_{N}k'}(0)
|{\bf p}_{\pi};in>
u({\bf p}_N)-
e_{\pi}<out;{\bf p'}_{{ p}_{\pi}}{\bf p'}_{N}|j_{\bf p_{\pi}k'}(0)
|{\bf p}_{N};in>,
\eqno(2.17a)$$

$${\cal E}_{\gamma'\pi' N'-\pi N}^{\mu}=
-\Biggl[{\overline u}({\bf p'}_N)\gamma^{\mu}
{{\gamma_{\nu} (p'_N+k')^{\nu}+m_N }\over{
2p'_Nk'}}e_{N'}
<out;{\bf p'}_{\pi}|J_{\bf p'_{N}k'}(0)|{\bf p}_{\pi}{\bf p}_N;in>$$
$$+(2{p'}_{\pi}+k')^{\mu}{{e_{\pi'}}
\over{ {2p'_{\pi}k' }} }
<out;{\bf p'}_{N}|j_{\bf p'_{\pi}k'}(0)|{\bf p}_{\pi}{\bf p}_N;in>$$
$$-e_N
<out;{\bf p'}_{\pi}{\bf p'}_{N}|{\overline J}_{\bf p'_{N}k'}(0)
|{\bf p}_{\pi};in>
{ {\gamma_{\nu} (p_N-k')^{\nu}+m_N}
\over{2p_Nk'} }\gamma^{\mu}
u({\bf p}_N)$$
$$-<out;{\bf p'}_{\pi}{\bf p'}_{N}|j_{\bf p_{\pi}k'}(0)|{\bf p}_{N};in>
{{e_{\pi}}\over{2p_{\pi}k'} }
(2p_{\pi}-k')^{\mu}\Biggr]\eqno(2.17b)$$

The relations (2.16) and (2.17a,b) 
have the same form  as (2.6) and (2.8a,b) in the formulation 
without quarks \cite{MFNEW}. The only differences are
in the source operators of the nucleons and pions. 
The off shell $\pi N$ amplitudes in (2.17a,b) contains   
the nonlocal source operators (2.13a,b) of composite 
particles which in contrast to the local
sources  $J(x)=(i\gamma_{\nu}\partial/\partial x_{\nu}-m_N)\Psi(x)$
and $j_{\pi}(x)=(\Box_x+m_{\pi}^2)\Phi(x)$
depends on the four moments of the composed particle 
and on ${\bf k'}$. Consequently, 
the off mass shell $\pi N$
amplitudes in (2.17a,b) have an additional dependence on the
Mandelstam variables.

The Ward-Takahashi identity (2.16) presents the general scheme of
current conservation for the $\pi N$ bremsstrahlung reaction with the 
composed on mass shell pions and nucleons. According to this scheme 
it is necessary to 
find a special part of the  internal particle radiation amplitude
${\cal I}_{\gamma'\pi' N'-\pi N}^{\mu}$ 
which insures current conservation because 
$$k'_{\mu}{\cal I}_{\gamma'\pi' N'-\pi N}^{\mu}={\cal B}_{\gamma'\pi' N'-\pi N}
\eqno(2.18a)$$

and consequently

$$k'_{\mu}
<out;{\bf p'}_{N}{\bf p'}_{\pi}|{\cal J}^{\mu}(0)|{\bf p}_{\pi}{\bf p}_N;in>
=k'_{\mu}
\biggl({\cal E}^{\mu}_{\gamma'\pi'N'-\pi N}
+{\cal I}_{\gamma'\pi' N'-\pi N}^{\mu}\biggr)=0.\eqno(2.18b)$$

An example of such an internal particle radiation amplitude is the
intermediate on mass shell $\Delta$ radiation amplitude depicted in
Fig. 3 \cite{MFNEW}. 
The $\Delta$ radiation amplitude in Fig. 3 
${\cal I}_{\gamma'\pi' N'-\pi N}^{\mu}(\Delta-\gamma\Delta)$
does not satisfy current conservation separately.
${\cal I}_{\gamma'\pi' N'-\pi N}^{\mu}(\Delta-\gamma\Delta)$ satisfy
current conservation together with
$({\cal E_L}^{3/2})_{\gamma'\pi' N'-\pi
  N}^{\mu}({\Delta}-\gamma\Delta)$
which is extracted from ${\cal E}_{\gamma'\pi' N'-\pi N}^{\mu}$
(2.17b)
in Fig.2 after the set of the decompositions.

$$k'_{\mu}({{\cal E_L}^{3/2}})^{\mu}_{\gamma'\pi'N'-\pi N}(\Delta-\gamma\Delta)
=-{k'}_{\mu}{\cal I}_{\gamma'\pi' N'-\pi N}^{\mu}(\Delta-\gamma\Delta)=
-{\cal B}^{3/2}_{\pi' N'-\pi N}(\Delta-\gamma\Delta),\eqno(2.19)$$

where the lower index $ _{\cal L}$ and the upper index $ ^{3/2}$ denotes 
the longitudinal and the spin-isospin  $(3/2,3/2)$ part of the corresponding 
expressions. From the same structure of 
$({\cal E_L}^{3/2})_{\gamma'\pi' N'-\pi N}^{\mu}({\Delta}-\gamma\Delta)$ 
and 
$({\cal I}^{3/2})_{\gamma'\pi' N'-\pi N}^{\mu}({\Delta}-\gamma\Delta)$
follows

$$({\cal E_L}^{3/2})^{\mu}_{\gamma'\pi'N'-\pi N}(\Delta-\gamma\Delta)=-
{\cal I}^{\mu}_{\gamma'\pi'N'-\pi N}(\Delta-\gamma\Delta)\eqno(2.20)$$
which allows to equate the  
$\Delta-\gamma\Delta$ vertex functions in 
${\cal I}^{\mu}_{\gamma'\pi'N'-\pi N}(\Delta-\gamma\Delta)$
and in $({\cal E_L}^{3/2})^{\mu}_{\gamma'\pi'N'-\pi N}(\Delta-\gamma\Delta)$

$$G_{C0}({k'},s,s')=
-2M_{\Delta}
\biggl[
{{|{\bf k'}|}\over {s-s'}}
-{ {{P}^o_{\Delta}(s)-{P'}^o_{\Delta}(s')}\over{s-s'}}
\biggr]$$
$$\Bigl[{\rm g}_{\pi' N'-\Delta'}(s',k')\Bigr]^{-1}
\biggl(e_N{{{\cal R}_{N'}+{\cal R}_N}\over 2}
+e_{\pi}{{{\cal R}_{\pi'}+{\cal R}_{\pi}}\over 2}\biggr)
\Bigl[{\rm g}_{\Delta-\pi N}(s)\Bigr]^{-1},
\eqno(2.21a)$$

$$G_{M1}({k'}_{\Delta},s,s')=
-2M_{\Delta}\biggl[
{{|{\bf k'}|}\over {s-s'}}
-{ {{P}^o_{\Delta}(s)-{P'}^o_{\Delta}(s')}\over{s-s'}}
\biggr]
\Bigl[{\rm g}_{\pi' N'-\Delta'}(s',k')\Bigr]^{-1}
\biggl(\mu_N {{{\cal R}_{N'}+{\cal R}_N}\over 2} \biggr)
\Bigl[{\rm g}_{\Delta-\pi N}(s)\Bigr]^{-1},
\eqno(2.21b)$$

where 
$G_C$ and $G_{M1}$ denote the electric and 
 magnetic  dipole form factors of the $\Delta$'s,
$k'_{\Delta}=P_{\Delta}-P'_{\Delta}$ and  we have
considered the $\pi N$ bremsstrahlung reactions with only 
$e_N=e_{N'}$ and  $\mu_N=\mu_{N'}$.
Equations (2.21a,b) present a relationship between
  $G_{C0}({k'}_{\Delta}^2,s,s')$,
$G_{M1}({k'}_{\Delta}^2,s,s')$ and the residues of the $\pi N$
amplitudes ${\cal  R}$ (see (A.9a,b,c,d) in \cite{MFNEW}).

At the threshold $k'=0$ one obtains  the same 
model-independent relation of 
the magnetic dipole  moments of the $\Delta^+$ and 
$\Delta^{++}$ resonances as in \cite{MFNEW}
$\mu_{\Delta^+}={ {M_{\Delta}}\over {m_p}} \mu_p$ 
and $\mu_{\Delta^{++}}={3\over 2}\mu_{\Delta^+}$.

\begin{center}
                  {\bf 3. Magnetic dipole moments of the  $\Delta^o$
                    and $\Delta^-$ resonances. }
\end{center}
\medskip

The external particle radiation amplitude 
${\cal E}_{\gamma'\pi' N'-\pi N}^{\mu}$ (2.17b) can be replaced 
by the one on mass shell particle exchange amplitude

$${\cal E}_{\gamma'\pi' N'-\pi N}^{\mu}(N)=-\Biggl[
(2{p'}_{\pi}+k')^{\mu}{ {e_{\pi'}}
\over{ {2p'_{\pi}k' }} }
<out;{\bf p'}_{N}|j_{\bf p'_{\pi'}k'}(0)|{\bf p}_{\pi}{\bf p}_N;in>
$$
$$ +{{
{\overline u}({\bf p'}_N)\Bigl[(2p'_N+k')^{\mu}\tau_+
-i\mu_{N'}\sigma^{\mu\nu}k'_{\nu}\Bigr] }\over{ 2p'_Nk'} }
u({\bf p_N'+k'}){\overline u}({\bf p_N'+k'})
 e_{N'}<out;{\bf p'}_{\pi}|J_{\bf p'_{N}k'}(0)|{\bf p}_{\pi}{\bf p}_N;in>
$$
$$
-e_N<out;{\bf p'}_{\pi}{\bf p'}_{N}|{\overline J}_{\bf p_{N}k'}(0)
|{\bf p}_{\pi};in>u({\bf p_N-k'}){\overline u}({\bf p_N-k'})
{ {\Bigl[(2p_N-k')^{\mu}-i\mu_N\sigma^{\mu\nu}k'_{\nu}
\Bigr]u({\bf p}_N) }\over{2p_Nk'} }$$
$$-<out;{\bf p'}_{\pi}{\bf p'}_{N}|j_{\bf p_{\pi'}k'}(0)|{\bf p}_{N};in>
{{e_{\pi}}\over{2p_{\pi}k'} }
(2p_{\pi}-k')^{\mu} \Biggr],\eqno(3.1a)$$
where  the antiparticle contributions are separated  as it was done
in \cite{MFNEW}. 
Thus starting from the generalized $S$-matrix 
reduction formulas (2.8a,b) for the 
$\pi N$ radiation amplitude with composite pions and nucleons one 
obtains again the modified Ward-Takahashi identity

$$
{k'}_{\mu}
<out;{\bf p'}_{N}{\bf p}_{\pi'}|{\cal J}^{\mu}(0)|{\bf p}_{\pi}{\bf p}_N;in>=
{\cal B}_{\pi' N'-\pi N}+{k'}_{\mu}{\cal E}_{\gamma'\pi' N'-\pi N}^{\mu}(N)
=0,\eqno(3.1b)$$

with the external particle radiation amplitude (3.1a),
where  $e_p=e$,  $ \mu_p=1$ for the protons
and $e_n=0$, $ \mu_n=0$ for the neutrons. 

In order to take into account the anomalous magnetic moments of nucleons
it is necessary to consider the loop corrections of the
$\gamma NN$ vertex.  The corresponding contributions can be extracted from 
the transverse part of ${\cal E}_{\gamma'\pi' N'-\pi N}^{\mu}(N)$. 
Then one obtains again the expression (3.1a) 
with $\mu_p=2.79\mu_B$ for protons and
$\mu_n=-1.91\mu_B$ for neutrons in the units of the 
nuclear magneton $\mu_B=e/2m_p$.

The  external particle radiation part of the $\pi N$ 
bremstrahlung amplitude with the complete 
$\gamma' NN$ and $\gamma'\pi\pi$ vertices is depicted  in Fig.2.   
At the threshold $k'=0$ (2.21a,b)  presents the exact
relationship between $e_{\Delta}$, $\mu_{\Delta}$
and the $\Delta$ pole residues 
 ${\cal R}$ 
of the off shell $\pi N$  amplitudes
(see (A.9a,b,c,d) in \cite{MFNEW})

$$e_{\Delta}=
-\Biggl[{\cal N}(s)
\Bigl[{\rm g}_{\pi' N'-\Delta'}(s',k')\Bigr]^{-1}
\biggl(e_N{{{\cal R}_{N'}+{\cal R}_N}\over 2}
+e_{\pi}{{{\cal R}_{\pi'}+{\cal R}_{\pi}}\over 2}\biggr)
\Bigl[{\rm g}_{\Delta-\pi N}(s,k')\Bigr]^{-1}
\Biggr]^{k'= 0}_{\sqrt{s'}=\sqrt{s}= M_{\Delta}},
\eqno(3.2a)$$

$$\mu_{\Delta}=-\Biggl[{\cal N}(s)
\Bigl[{\rm g}_{\pi' N'-\Delta'}(s',k')\Bigr]^{-1}
\biggl(\mu_N {{{\cal R}_{N'}+{\cal R}_N}\over 2} \biggr)
\Bigl[{\rm g}_{\Delta-\pi N}(s,k')\Bigr]^{-1}
\Biggr]^{k'= 0}_{\sqrt{s'}=\sqrt{s}= M_{\Delta}},
\eqno(3.2b)$$

where 
${\cal N}(s)=1/(d{\sqrt{s}}/dk')-d{P}^o_{\Delta}(s)/d{\sqrt{s}} $ and
${\rm g}_{\Delta-\pi N}$ and ${\rm g}_{\pi' N'-\Delta'}$ denotes the 
$\Delta-\pi N$ form factors.

Now we assume that the charge of the neutron is an auxiliary 
parameter $e_{n}$ which will be fixed in the finally relations as
$e_{n}=0$. 
Then for the $\pi N$ bremsstrahlung with the intermediate $\Delta^o$   
we have

$$e_{\Delta^o}=
-e_n\Biggl[{\cal N}(s)
\Bigl[{\rm g}_{\pi' N'-\Delta'}(s',k')\Bigr]^{-1}
{{{\cal R}_{n'}+{\cal R}_n}\over 2}
\Bigl[{\rm g}_{\Delta-\pi N}(s,k')\Bigr]^{-1}
\Biggr]^{k'= 0}_{\sqrt{s'}=\sqrt{s,k'}= M_{\Delta}}.
\Biggr]^{k'= 0}_{\sqrt{s'}=\sqrt{s}= M_{\Delta}}
\eqno(3.3a)$$
But $e_{\Delta^o}=e_n$ for the reaction $\pi^o n\to\gamma'{\pi^o}'n'$.
The cancellation of $e_n$ from
both sides of (3.3a) gives the normalization condition 
for ${\cal R}_n$

$$1=-\Biggl[{\cal N}(s)
\Bigl[{\rm g}_{\pi' N'-\Delta'}(s',k')\Bigr]^{-1}
{{{\cal R}_{n'}+{\cal R}_n}\over 2}
\Bigl[{\rm g}_{\Delta-\pi N}(s,k')\Bigr]^{-1}
\Biggr]^{k'= 0}_{\sqrt{s'}=\sqrt{s}= M_{\Delta}}.
\eqno(3.3b)$$

Substituting (3.3b) into (3.2b) we obtain
 
$$\mu_{\Delta^o}=
\mu_n{ {M_{\Delta} }\over {m_p} },\eqno(3.4)$$
where the different units of $\mu_{\Delta}$ and $\mu_N$
generates the factor ${ {M_{\Delta} }/ {m_p} }$.
The isospin symmetry between the $\pi N$ amplitudes of the
reactions $\pi^o n\to \pi^o n$ and $\pi^- n\to\pi^- n$ 
in (3.2b) allows  to estimate $\mu_{\Delta^-}$

$$\mu_{\Delta^{-}}={3\over 2}\mu_{\Delta^o}={3\over 2}
\mu_n{ {M_{\Delta} }\over {m_p} }
\eqno(3.5)$$


\newpage

\begin{center}
                  {\bf 4. Conclusion}
\end{center}
\medskip

The main result of this paper
is that the magnetic dipole  moments of the $\Delta$ resonances
are  the same
in the quantum field theories with and without 
quark degrees of freedom. 
This follows from the same structure of  
 current conservation   for the on shell 
$\pi N$ radiation amplitudes in the formulations with and without 
quark degrees of freedom,
and the simple relationship  between the 
magnetic dipole  moments of the $\Delta$'s and the anomalous magnetic 
moment of the nucleons
$\mu_{\Delta}=M_{\Delta}/m_p\ \mu_{N}$.
According to this formula the 
$\mu_{\Delta}$  are dependent only on the anomalous magnetic moment of
 the nucleons
$\mu_N$ and the $\Delta$ resonance pole position 
$M_{\Delta}=1232\ MeV$.
The present approach allows  to connect analytically 
the  electric and the  magnetic dipole form factors $G_C$ and $G_{M1}$
with the $\Delta$ pole residues ${\cal R}$ of the off shell $\pi N$ 
amplitudes according to (2.21a,b). 
The $\Delta$ pole residues ${\cal R}$ as well as the 
source operators (2.13a,b) and (A.6a,b) are different
in the different models.
Correspondingly, the
dependence on $k'$ of the off shell $\pi N$ amplitudes is different
and model-dependent. But at threshold $k'=0$ the expressions for
 $e_{\Delta}$, $\mu_{\Delta }$ (3.2a,b) as well as 
the normalization condition (3.3b) are the same for any model
with the fixed charge of the particles. 
Therefore, the formulas (3.4) and (3.5) for the 
 magnetic dipole moment of the $\Delta$'s are unique and 
model-independent.

\vspace{0.15cm}

\centerline{\bf Table  1}

\vspace{0.15cm}

\begin{center}
{\em Magnetic moments of $\Delta^o$ and $\Delta^{-}$
 in nuclear magneton $\mu_B={e/{2m_p}}$. 
}   
\end{center}
\hspace{-0.75cm}

\begin{center}
\begin{tabular}{|c|c|c|c|c|c|c|c|} \hline\hline
{\sc models}&  {\em This}                   &  {\em $SU(6)$ }
            &  {\em Skyrme }                &
\\
            &      {\em  work }             & {\em and Bag } 
            &                               &  {\em quark}      

\\    \hline

                       &                             & 0. \cite{Beg,Georgi}
                       &                             & 0.\cite{Buch97}         
                                 
\\

${ \mu_{\Delta^{o}}}$ & -2.504                       &  0.\cite{Pais} 
                      & -1.33$\sim$-0.19\cite{Acu98} & 0.375\cite{Kim98}

\\
                       &                     & 0.\cite{Brown}
                       &                     &-0.3$\sim$0.\cite{Lin98}

\\                     &                             &0.\cite{Kriv} 
                       &                             &

\\    \hline

                       &                             &  -2.79 \cite{Beg,Georgi}
                       &                             &-3.49\cite{Buch97}         
                                 
\\

${ \mu_{\Delta^{-}}}$ & -3.759                    &  -2.13\cite{Pais} 
                & -5.62$\sim$-2.38\cite{Acu98}    & -2.1\cite{Kim98}

\\
                       &                      &-2.20-2.45\cite{Brown}
                       &                      & -2.72-3.06\cite{Lin98}         

\\                     &                             &-3.27\cite{Kriv} 
                       &                             &

\\ \hline \hline

\end{tabular}
\end{center}

\vspace{0.45cm}


The present
relations  for $\mu_{\Delta}$ requires proportionality of $\mu_{\Delta}$
and the anomalous magnetic moment of the nucleon $\mu_{N}$. 
In particular, $\mu_{\Delta^o}$ and $\mu_{\Delta^-}$
are determined via the anomalous magnetic moment of the neutron.
A comparison our numerical values  
for $\mu_{\Delta^o}$ and $\mu_{\Delta^-}$ with the calculations of 
other authors is given in Table 1.
In the $SU(6)$ symmetry quark models 
and their modifications \cite{Beg}-\cite{Kriv}
  $\mu_{\Delta}$ is proportional to the charge of the $\Delta$.
Therefore in these models\cite{Beg,Georgi,Pais,Brown,Kriv}
 $\mu_{\Delta^o}=0$ and 
$\mu_{\Delta^-}=-\mu_{\Delta^+}$ and  
$\mu_{\Delta^+}=1/2\ \mu_{\Delta^{++}}$.
This property is preserved  in the constituent quark model \cite{Buch97}.
But it is slightly broken in the Skyrme model \cite{Acu98},
chiral quark model \cite{Lin98}, chiral quark-soliton 
model \cite{Kim98} and effective quark model. 
The crucial difference between our result and the other estimations is
in $\mu_{\Delta^o}$ which is larger  than the predictions of
other authors.

\vspace{7mm}


\begin{center}
                  {\bf Appendix A: Alternative 
field operators of composite
particles}
\end{center}

\vspace{7mm}

The source operators $J_{p'_{N},k'}(Y)$ (2.13a) and $j_{{ p'}_{\pi},k'}(X)$
(2.13b) can be constructed in the independent over the quark charges $e_j$
form. For this aim one can introduce  other  quark cluster operators

$$\Psi_{p_N}(Y)={1\over 6}
\int d^4r_3d^4r_{1,2}\Biggl\{$$
$$\biggl[{\widetilde \chi}^{\dag}_{p_N}(y_1,y_2.y_3)|_{Y=0}
{\sf T}\biggl( q_1(y_1)q_2(y_2)q_3(y_3) \biggr)
+{\widetilde \chi}^{\dag}_{p_N}(y_2,y_1.y_3)|_{Y=0}
{\sf T}\biggl( q_1(y_2)q_2(y_1)q_3(y_3) \biggr)\biggr]$$
$$+\biggl[
{\widetilde \chi}^{\dag}_{p_N}(y_3,y_2.y_1))|_{Y=0}
{\sf T}\biggl( q_1(y_3)q_2(y_2)q_3(y_1) \biggr)
+{\widetilde \chi}^{\dag}_{p_N}(y_2,y_3.y_1))|_{Y=0}
{\sf T}\biggl( q_1(y_2)q_2(y_3)q_3(y_1) \biggr)\biggr]$$
$$+\biggl[
{\widetilde \chi}^{\dag}_{p_N}(y_1,y_3.y_2))|_{Y=0}
{\sf T}\biggl( q_1(y_1)q_2(y_3)q_3(y_2) \biggr)
+{\widetilde \chi}^{\dag}_{p_N}(y_1,y_1.y_2)|_{Y=0}
{\sf T}\biggl( q_1(y_3)q_2(y_1)q_3(y_2) \biggr)\biggr]
\Biggr\}\eqno(A.1)$$

and

$$\Phi_{{ p}_{\pi}}(X)={1\over 2}\Biggl\{\int d^4\rho_{1,2}
{\widetilde \phi}^+_{p_{\pi}}(X=0,\rho_{1,2}){\sf T}\biggl(q_i(x_1)
{\overline q}_i(x_2)\biggr)
+\int d^4\rho_{1,2}
{\widetilde \phi}^+_{p_{\pi}}(X=0,-\rho_{1,2}){\sf T}\biggl(q_i(x_2)
{\overline q}_i(x_1)\biggr)
\Biggr\},\eqno(A.2)$$

Unlike to  (2.1) and (2.4a), the field operators (A.1) and
(A.2) contains all transpositions of the integration variables
$y_1$, $y_2$, $y_3$ and $x_1$, $x_2$.
Therefore,  instead of (2.8b) we obtain

$${k'}_{\mu}{\sc G}^{\mu}=i\int d^4[y']
d^4[x']d^4[y]d^4[x]
{\cal P}_{x'_1x'_2}{\cal P}_{x_1x_2}
{\cal P}_{y'_1y'_2y'_3}{\cal P}_{y_1y_2y_3}
e^{ik'z}d^4z
{\widetilde \chi}^+_{p'_N}(y'_1,y'_2,y'_3)
{\widetilde \phi}^+_{p'_{\pi}}(x'_1,x'_2)
{{\partial}\over{\partial z^{\mu}} }$$
$$<0|{\sf T}\biggl(
{\sf T}\Bigl(q_1(y'_1)q_2(y'_2)q_3(y'_3)\Bigr)
{\sf T}\Bigl(q_1(x'_1){\overline q}_2(x'_2)\Bigr)
{\cal J}^{\mu}(z)
{\sf T}\Bigl({\overline q}_1(y_1){\overline q}_2(y_2){\overline q}_3(y_3)\Bigr)
{\sf T}\Bigl({\overline q}_1(x_1)q_2(x_2)\Bigr)\biggr)|0>$$
$$\chi_{p_N}(y_1,y_2,y_3)\phi_{p_{\pi}}(x_1,x_2)\eqno(A.3)$$

where $d^4[y]\equiv1/6\Bigl\{[d^4y_1d^4y_2d^4y_3+d^4y_2d^4y_1d^4y_3]
+[d^4y_3d^4y_1d^4y_2+d^4y_1d^4y_3d^4y_2]+
[d^4y_1d^4y_2d^4y_3+d^4y_2d^4y_3d^4y_1]\Bigr\}$,
$d^4[x]\equiv 1/2\Bigl\{d^4x_1d^4x_2+d^4x_2d^4x_1\Bigr\}$
and ${\cal P}_{x_1x_2}$ and ${\cal P}_{y_1y_2y_3}$ are defined 
via  transposition operator ${\sf P}_{x_1x_2}$ of the variables 
$x_1$ and $x_2$ as 

$${\cal P}_{x_1x_2}={{1+{\sf P}_{x_1x_2}}\over 2}
;\ \ \  {\cal P}_{y_1y_2y_3}={{{\sf P}_{y_1y_2}+{\sf P}_{y_2y_3}
+{\sf P}_{y_3y_1}
}\over 3}
\eqno(A.4)$$

The symmetry over the rearrangement of the
integration variables in (A.3) and 
$\int dx_1dx_2e^{ik'x_1}[f(x_1,x_2)+f(x_2,x_1)]=
\int dx_2dx_1e^{ik'x_2}[f(x_2,x_1)+f(x_1,x_2)]$ 
allows to modify (2.11a) as

$${k'}_{\mu}A_{\gamma'\pi' N'-\pi N}^{\mu}=
-ie_{N'}{\overline u}({\bf p'}_N)(\gamma_{\nu} {p'_N}^{\nu}-m_N)
\int d^4[y']{\cal P}_{y'_1y'_2y'_3}
{\widetilde \chi}^+_{p'_N}(y'_1,y'_2,y'_3)$$
$${{e^{ik'y'_1}+e^{ik'y'_2}+e^{ik'y'_3}}\over 3}<out;{\bf p'}_{\pi}|
{\sf T}\Bigl(q_1(y'_1)q_2(y'_2)q_3(y'_3)\Bigr)
|{\bf p}_{\pi}{\bf p}_N;in>$$
$$-ie_{\pi'}({p'_{\pi}}^2-m_{\pi}^2)\int d^4[x']{\cal P}_{x'_1x'_2}
{\widetilde \phi}^+_{p'_{\pi}}(x'_1,x'_2)
{{e^{ik'x'_1}+e^{ik'x'_2}}\over 2}
<out;{\bf p'}_{N}|{\sf T}\Bigl(q_1(x'_1){\overline q}_2(x'_2)\Bigr)
|{\bf p}_{\pi}{\bf p}_N;in>$$
$$+ie_{N}\int d^4[y]{\cal P}_{y_1y_2y_3}
{{e^{ik'y_1}+e^{ik'y_2}+e^{ik'y_3}}\over 3}
<out;{\bf p'}_{N}{\bf p'}_{\pi}
|{\sf T}\Bigl({\overline q}_1(y_1){\overline q}_2(y_2){\overline q}_3(y_3)
\Bigr)|{\bf p}_{\pi};in>$$
$$\chi_{p_N}(y_1,y_2,y_3)
(\gamma_{\nu} {p_N}^{\nu}-m_N)({p_{\pi}}^2-m_{\pi}^2)u({\bf p}_N)$$
$$+ie_{\pi}({p'_{\pi}}^2-m_{\pi}^2)\int d^4[x]{\cal P}_{x_1x_2}
{{e^{ik'x_1}+e^{ik'x_2}}\over 2}
<out;{\bf p'}_{N}{\bf p'}_{\pi}|{\sf T}\Bigl( {\overline q}_1(x_1)q_2(x_2)
\Bigl)|{\bf p}_{N};in>\phi_{p_{\pi}}(x_1,x_2)
,\eqno(A.5)$$

Consequently,  after integration over 
 $X'$, $X$, $Y'$ and $Y$ we obtain again (2.12) with
independent on the quark charges $e_j$  source operators 

$$J_{p'_{N},k'}(Y)=
\bigl(i\gamma_{\sigma}\nabla_Y^{\sigma}-m_{N}\bigr)\Psi_{{\bf p}_{N},k'}(Y)
;\ \ \Psi_{{\bf p'}_{N},k'}(Y')=
\int d^4r'_3d^4r'_{1,2}
{\widetilde \chi}^+_{p'_N}(Y'=0,r'_{1.2},r'_3)$$
$${{e^{ik'y'_1}+e^{ik'y'_2}+e^{ik'y'_3}}\over 3}|_{Y'=0}
{\sf T}\Bigl(q_1(y'_1)q_2(y'_2)q_3(y'_3)\Bigr)\eqno(A.6a)$$

$$ j_{{ p'}_{\pi},k'}(X)=
\bigl(\Box_X+m_{\pi}^2\bigr)\phi_{ p_{\pi}}(X)
;\ \ \phi_{ p_{\pi},k'}(X')=
\int d^4\rho'_{1,2}
{\widetilde \phi}^+_{p'_{\pi}}(X'=0,\rho'_{1,2})$$
$${{e^{ik'x'_1}+e^{ik'x'_2}}\over 2}|_{X'=0}  
{\sf T}\Bigl(q_1(x'_1){\overline q}_2(x'_2)\Bigr).
\eqno(A.6b)$$



\end{document}